\documentclass[12pt,preprint]{aastex}

\shorttitle{M31 and M33 Upper HR Diagram}
\shortauthors{Humphreys et al. }

\begin{document}

\title{Luminous and Variable Stars in M31 and M33 V. The Upper HR Diagram}

\author{
Roberta M. Humphreys\altaffilmark{1}, 
Kris Davidson\altaffilmark{1},
David Hahn\altaffilmark{1},
John C. Martin\altaffilmark{2} and 
Kerstin Weis\altaffilmark{3}
}

\altaffiltext{1}
{Minnesota Institute for Astrophysics, 116 Church St SE, University of Minnesota
, Minneapolis, MN 55455; roberta@umn.edu} 

\altaffiltext{2}
{Barber Observatory, University of Illinois, Springfield, IL, 62703}

\altaffiltext{3}
{Astronomical Institute, Ruhr-Universitaet Bochum, Germany}

\begin{abstract}
We present HR Diagrams for the massive star populations in M31 and M33  
including several different types of emission-line stars: the confirmed Luminous
Blue Variables (LBVs), candidate LBVs, B[e] supergiants and the warm hypergiants. We 
estimate their apparent temperatures and luminosities for 
comparison with their respective massive star populations and to evaluate the possible 
relationships of these different classes of evolved, massive  stars, and their evolutionary state. 
Several of the LBV candidates
lie near the LBV/S Dor instability strip which supports their classification. Most of the
B[e] supergiants, however,  are less luminous than the LBVs. Many are very dusty with the infrared flux contributing one-third or more to their total flux. They are also relatively isolated from  other luminous OB stars. Overall, their spatial distribution suggests a more evolved state. Some may be post-RSGs like the warm hypergiants, and there  
may be more than one path to becoming a B[e] star. 
There are sufficient differences in the spectra, luminosities, spatial distribution, and the presence or lack of dust between the LBVs and B[e] supergiants to conclude that one group does not evolve into the other.  
\end{abstract} 


\section{Introduction -- The Complex Upper HR Diagram }

The upper HR Diagram is populated with some of the most interesting and challenging stars with respect to their physics, evolution and eventual 
fate. Massive stars  are distinguished by their relatively short lifetimes, mass loss and their eventual fate as supernovae, 
neutron stars and black holes. In addition to stellar winds and mass loss,
many of them show evidence for periods of enhanced mass loss, such as the Luminous Blues Variables (LBVs) and the warm and cool hypergiants with resolved ejecta. In addition, Wolf-Rayet stars of various types, Oe and Of stars, the B[e] supergiants, and the Fe II emission line stars occupy the same parts of the HR Diagram.  Supernovae surveys have identified numerous non-terminal giant
eruptions, in which the object greatly increases its total luminosity possibly expelling several solar masses. Some of these events are confused with true SNe and thus have been  called ``supernova impostors'' \citep{VanDyk05,vandyk}. Very little is known about their progenitors, although several have been 
identified with likely massive stars.  The nature of their instability is unknown, but proximity to the 
 Eddington Limit may be crucial.  The pre-eruption stars very likely 
  have  $L/M \sim 0.5 \, (L/M)_\mathrm{Edd}$  like LBVs \citep{RMH16}, 
   but there is no proof that they are indeed classical LBVs/S Doradus variables.  Most of the other categories mentioned above may have lower 
     $L/M$, but rotation can magnify its effective value in 
      B[e] stars, for example. 

Thus we observe a complex upper HR diagram with several types of stars 
not only experiencing continuous mass loss, but also  high mass loss events. Some of these  classes of stars may be related to each other and may represent stars of similar mass
but in different 
stages in their evolution as they shed mass. Complicating our understanding 
are recent results suggesting that the most massive stars may not actually 
end their lives as supernovae. Smartt (2009, 2015) has suggested an upper mass 
limit of $\approx$ 18 M$\odot$ for the red supergiant progenitors of 
the Type II SNe, while Jennings et al.(2014)  find
a lack of massive progenitors in M31 and M33 and suggest an upper mass 
limit of 35-45 M$\odot$ for the Type II SNe. 

An improved census of the  upper HR Diagram , including the most 
luminous evolved stars, the LBVs, the hot emission line stars, and the 
warm and cool hypergiants is  needed for a more complete  
picture of the pre-terminal stages of very massive stars.  

In this series of papers  we have described the results of a spectroscopic survey of 
luminous and variable stars in the nearby spirals M31 and M33. In Paper I \citep{RMH13}
we discussed a small group of very luminous intermediate temperature supergiants, the warm hypergiants, 
and showed that they were likely post-red supergiants. In Paper II \citep{RMH14}, we reviewed  the spectral characteristics,  spectral energy distributions (SEDS), circumstellar
ejecta, and  mass loss for 82 luminous and variable stars including the 
confirmed LBVs, candidate LBVs, and other emission line stars. 
Many of these stars have circumstellar dust including several of 
the Fe II emission line stars, but found that the confirmed LBVs in M31 and 
M33 do not.  The confirmed LBVs also have 
relatively low wind speeds even in their hot, quiescent or visual minimum state 
compared to the 
B-type supergiants and Of/WN stars which they spectroscopically resemble.

\citet{Gordon} (Paper III) presented  a more comprehensive spectroscopic survey of the yellow supergiants. Based on spectroscopic 
evidence for mass loss and the presence of circumstellar dust in their SEDs, 
we conclude that $30-40\%$ of the yellow supergiants are likely in a post-red 
supergiant state. Comparison with evolutionary tracks shows that these 
mass-losing, post-RSGs have initial masses between 20 to 40 M$_{\odot}$ suggesting
that red supergiants in this mass range evolve back to warmer temperatures before their terminal state. 

In Paper IV \citep{RMH17}, we reported  spectroscopy of 132 additional luminous stars
and emission line objects including LBVs and  candidates, the B[e] supergiants, and the warm 
hypergiants.  Many of these stars are spectroscopically similar and are often confused with each other. We discussed their similarities and differences and  proposed  guidelines that can be used to help
distinguish them in future work.

In this final paper we present the upper HR Diagrams for M31 and M33 based 
on this work and that of Massey (2016). We determine the luminosities of the 
emission line stars from Papers II and  IV to place them on the HR Diagram 
for comparison with their
respective massive star populations and a discussion of their possible evolutionary state.
In the next section we briefly describe the observations used for what we call the representative  supergiant or massive star populations, the corrections for interstellar extinction, and 
their luminosities.  In \S {3}, we discuss the SEDs, circumstellar dust, interstellar extinction and the derivation of the luminosities and temperature estimates for the different types of emission line stars. In \S {4} we present the HR diagrams and compare the distribution of the emission line objects with the massive star 
populations in M31 and M33. In the final section we discuss the implications for their 
evolutionary states.  

\section{The Observational Data for the Supergiant Populations} 

We use the large data set published by \citet{Massey16}, together with our own spectral 
classifications from Papers II, III and IV, to create a catalog of luminous O,B and A type 
stars representative of the hot star populations in M31 and M33. We are careful to avoid
duplication, and  only stars with spectral types  are used. The WR stars are not included. All foreground stars and those labeled ``H ~II'' or ``double'' are omitted.  

In our previous work we have found that the interstellar extinction can vary considerably across the face of these galaxies, especially in M31. Since  many 
of these stars are also in nebulous and dusty regions, we determine the 
extinction for each star individually instead of assuming a mean extinction for
the host galaxy. We use its 
spectral type and corresponding intrinsic color together with the multi-color photometry from \citet{Massey06} and the standard extinction curve from 
\citet{Cardelli}  with $R = 3.2$. However, several stars in each galaxy 
have negative derived A$_{v}$'s. All except one are described as isolated (I) by \citet{Massey16}. We therefore assume that the anomalous A$_{v}$ values are due to photometric error or unresolved blends and replace them with our mean values, 0.62 $\pm$ 0.02 mag for M31 and 0.33 $\pm$ 0.01 mag for M33, determined from the other stars. The mean  A$_{v}$ for M33  is similar to the value adopted by \citet{Massey16}, but our mean for M31  is significantly higher than theirs.  

We also identified  13 stars with high A$_{v}$ values ($>$ 2 mag). High 
extinction in star forming regions in M31 and M33 is not necessarily erroneous. However most of these cases yielded physically implausible luminosities i.e. too high in M$_{v}$ and M$_{Bol}$. Eight were classed as extremely crowded (X) by \citet{Massey16}. We checked all 13 stars and confirm that these 8 were blended, unresolved images, or embedded in nebulosity. The remaining 5 were listed as isolated (I) meaning that their spectra and photometry are not contaminated by nearby stars or nebulosity. Their high extinction values are confirmed from neighboring stars and also from the neutral hydrogen density (see \S 3). 
Three have derived 
luminosities appropriate to their spectral types and are included in our catalog. The other two have luminosities $\sim$ 10$^{7}$ L$_{sun}$ which if correct imply  initial masses
above 250 M$_{\odot}$. These two stars are of special 
interest, deserve more attention, and are discussed in the Appendix. 

The absolute visual magnitudes are then derived from  distance moduli, 24.4 mag for M31 \citep{M31Ceph} and 24.5 mag for M33 \citep{M33Ceph}. 
The corresponding 
effective temperatures and bolometric corrections  are adopted  from  \citet{Martins} for the O-type stars and \citet{Flower} for the B and A-type supergiants. There are also numerous A-type supergiants  in Table 4 in our  Paper III. For those stars,  some of which have mass loss and circumstellar dust, we use our 
classifications and derived luminosities from  that paper. 

For the evolved yellow supergiants (YSGs) with F and G-type spectra  and 
the red supergiants (RSGs) we use the stars from  Figures 10 and 11  in our Paper III. Many  of these stars show an infrared excess in their
SEDs. This contribution to their luminosities  from  circumstellar dust is 
included in 
their bolometric luminosities by integrating their SEDs. 

Table 1 is a summary of the total number of stars of different spectral types 
used in our HR Diagrams for M31 and M33. These spectral type groups are not 
intended to be complete. They are used in this paper to  represent the massive star populations in their respective galaxies. The resulting HR Diagrams are shown in Figures 1 and 2. Although M31 and M33 have somewhat 
higher and lower metallicities, respectively,  than the Milky Way, we chose to use the evolutionary tracks from \citet{Ekstrom} for solar metallicity, because they cover a wide range of masses and include tracks with and without rotation.
In this paper they are used to provide an estimate of the likely mass range for some of the stars discussed later.   
In the next section, we discuss the luminosities and apparent temperatures  of the emission line stars for placement on the HR Diagrams for comparison with the representative  massive star populations shown here and with the evolutionary models. 

\begin{figure} 
\figurenum{1}
\epsscale{1.0}
\plotone{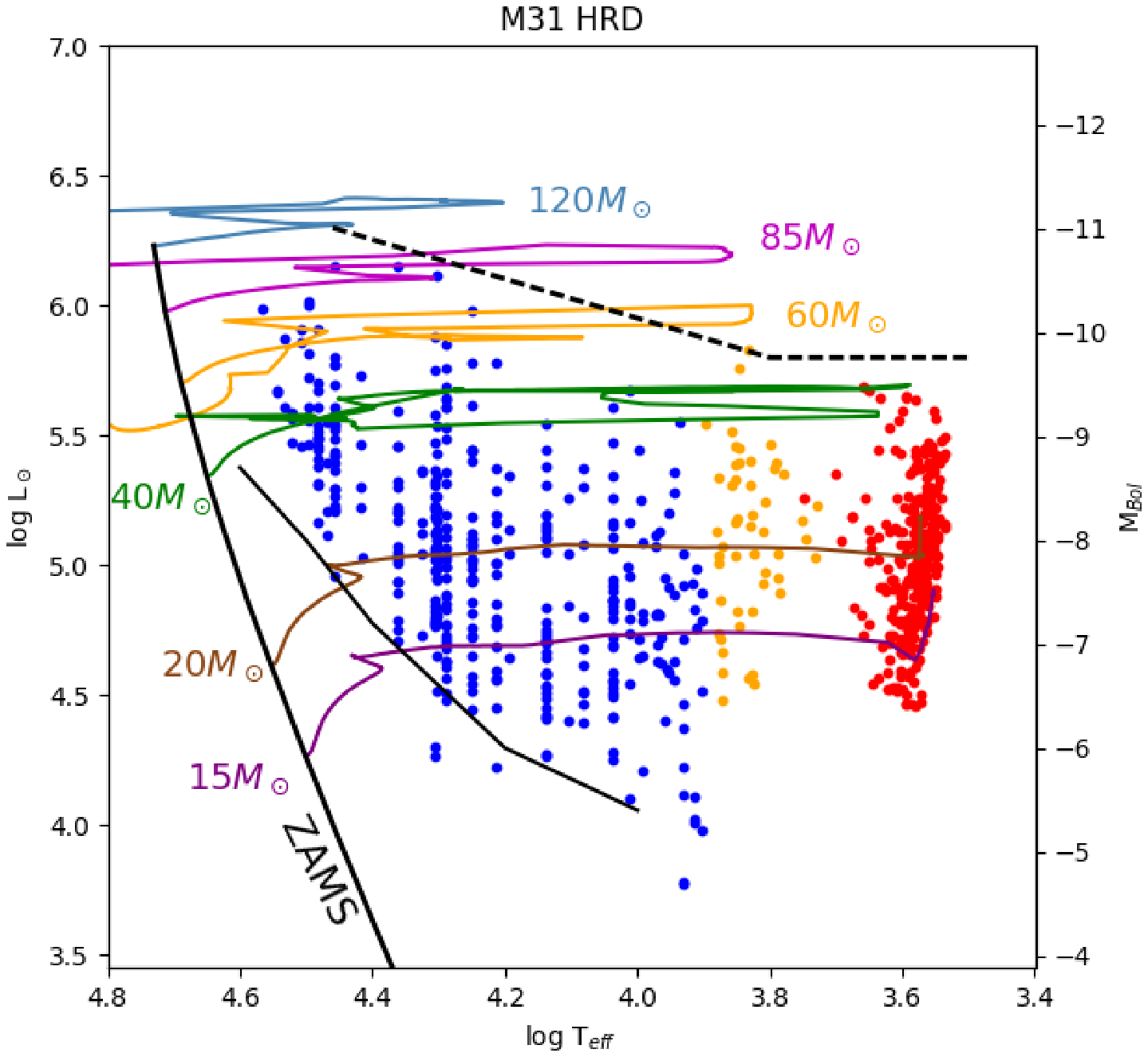}
\caption{The representative HR Diagram for the massive stars in M33. The OBA sta
rs are in blue, YSGS in yellow and RSGs in red. The evolutionary tracks and ZAMS
 are from \citet{Ekstrom} with mass loss, Milky Way metallicity and no rotation.
  The Humphreys-Davidson limit is shown as a dashed line. 
  The lower boundary on the HRD is set by Massey's selection of stars brighter 
  than M$_{V} = -5.5$ mag.} 
\end{figure}

\begin{figure}   
\figurenum{2}
\epsscale{1.0}
\plotone{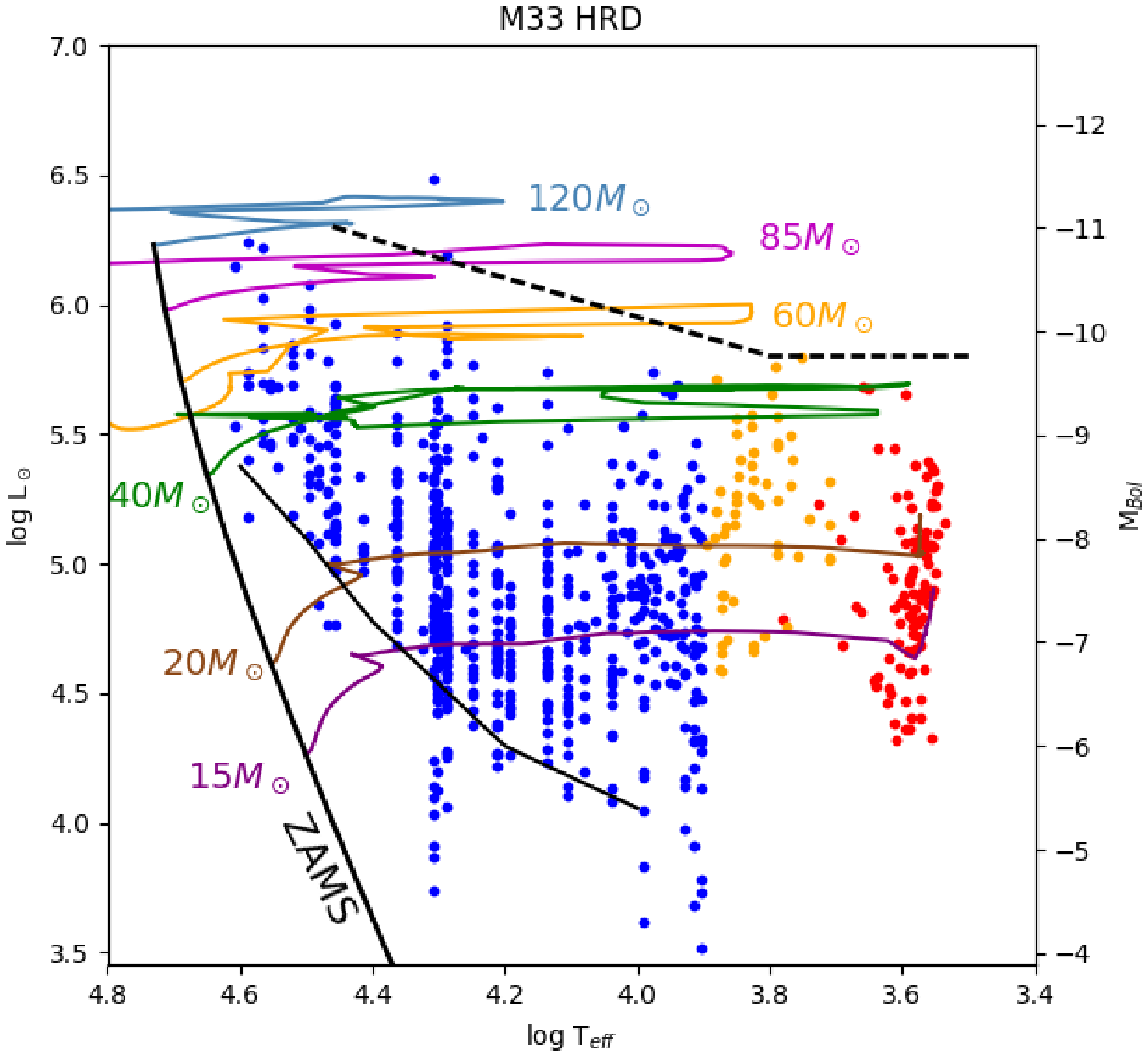}
\caption{The representative HR Diagram for the massive stars in M33.  The OBA st
ars are in blue, YSGS in yellow and RSGs in red. The evolutionary tracks and ZAM
S are from \citet{Ekstrom} with mass loss, Milky Way metallicity and no rotation
.. The
Humphreys-Davidson limit is shown as a dashed line. The lower boundary on the HR
D is set by Massey's selection of stars brighter than  M$_{V} = -5.5$ mag. The s
tar above the H-D limit is J013350.43+303833.8, marked as "extremely crowded" by
 Massey, and its photometry is probably blended with other stars.}  
\end{figure}

\section{LBVs and Other Emission Line Stars}

Our population of emission line stars are from Tables 5 and 6 in Paper IV and our list of candidate LBVs in Table 7 based on the guidelines  described in that paper.  
To place these stars with strong emission lines on the HR Diagram we must determine their intrinsic luminosities and estimate their surface temperatures. Their photometry and SEDs must first be corrected for interstellar extinction. This is an uncertain procedure for stars with strong emission lines because their broad-band colors in the blue-visual region cannot be safely used. Furthermore the lack of absorption lines, especially in the spectra of the B[e]sgs and Fe II emission line stars  (Paper IV), prevents accurate  spectral classification and adoption of the corresponding temperatures, and bolometric corrections.
Spectra of all of the stars discussed in this paper and the previous papers in this series are available at http://etacar.umn.edu/LuminousStars/M31M33. 

We follow the procedure described in Paper II. We first estimate the 
visual extinction, A$_{v}$, from  two independent methods, the Q-method for 
nearby OB-type stars, typically within 2 arcsec, based on their observed colors 
from  \citet{Massey06},  and the well-known relation between neutral hydrogen column density (N$_{H~I}$) and the color excess E$_{B-V}$ \citep{Knapp,Savage}. We define A$_{v}$ from N$_{H~I}$ as the foreground reddening (A$_{v} =$ 0.3 mag for M31 and 0.26 mag for M33) plus 1/2 A$_{v}$ from the N$_{H~I}$\footnote{We adopt half the extinction value because we do not know the exact locations of the stars along the line of sight with respect to the neutral hydrogen.}.   
 The results are summarized in Tables 2 and 3. We prefer the extinction estimates from neighboring stars when available because
of their proximity to the target stars, compared to the H I surveys which have spatial resolutions of 30$\arcsec$ for M31 \citep{Braun}  and 17$\arcsec$ for 
M33 \citep{Gratier}. The number of nearby stars used is given in parenthesis. The  
tables also include the adopted A$_{v}$ and the corresponding extinction-corrected 
absolute visual magnitudes, M$_{v}$. 
We also re-examined the extinction estimates for the stars in Paper II and they are included here. Differences in the parameters for the confirmed LBVs in 
Paper II and \citet{RMH16} are due to minor adjustments in the adopted 
extinction. 

To estimate their intrinsic luminosities and their surface temperatures, we integrate their extinction-corrected SEDs.  Many of these stars are also 
relatively hot. So to avoid understimating their total luminosity, the NUV and FUV fluxes, when available,    from the UIT survey \citep{Massey96}  
and the GALEX Nearby Galaxy Survey were included in the SEDs. The cross identification 
of our M33 stars with GALEX was aided by the catalog of UV sources in M33 by 
\citet{Mudd}. We used Galex View  to identify UV counterparts for the  
M31 stars. For some stars though, no UV counterpart was identified. We adopt the LMC average extinction curve  from \citet{KGordon} for the NUV and FUV fluxes. 
To illustrate the UV contribution to the total flux and how it constrains the 
Planck curve fit, we show  the  
SED for M33-013424.78  in Figure 3 with and without the UV flux.  A simple Planck curve is fit to the SEDs to estimate the surface  temperature. We use the Planck curve instead of trying to fit a model 
atmosphere, because the spectra of many of the stars are dominated by strong 
emission lines with a poorly defined continuum. More elaborate modeling is beyond the scope of this paper. A strong H$\alpha$ line can contribute significantly to the R-band photometry as can be seen in some of the SEDs. In those cases the R band photometry is not included in the fit.  The Planck fit to the SED 
is then integrated from 0.15$\mu$m to 0.8$\mu$m to determine the bolometric 
luminosities in the UV/blue-visual wavelength region.  

The primary source of uncertainty with this procedure is the
 adopted extinction correction  which  directly affects the derived
 temperatures and luminosities in the visual.
 The range in the A$_{v}$ values from the nearby stars
 is typically $\pm$ 0.2 mag. To estimate the expected
 uncertainty in the final adopted luminosities and temperatures, we use the
 range in the A$_{v}$ values from the nearby stars  for two different stars,
 one with a high luminosity and temperature (M33C-16364, 31000\arcdeg K, -9.4 M$_{Bol}$) and one of lower temperature and luminosity (M33C-2976, 17000\arcdeg K, -7.2 M$_{Bol}$). Varying the A$_{v}$ value by $\pm$ 0.2 mag for each gave a range in temperature of 
 about $\pm$ 16\% and $\pm$ 50\% in luminosity for the hotter star, 
 and $\pm$ 5\%   and $\pm$ 20\%  in temperature and luminosity, respectively
for the fainter star. These results will of course vary from star to star,
 but they give an indication of what uncertainties to expect for the stars'
 positions on the HR Diagrams.

\subsection{Circumstellar Dust}

The warm hypergiants and most of the B[e]sgs have significant excess radiation longwards of 1$\mu$m due to dust. The contribution to their total luminosities from their flux readiated by dust can be determined  
by integrating under a curve fit to the long wavelength data.

Figure 4 shows three examples, the warm hypergiant, M31-004444.52 and two  
B[e] supergiants, M31-004415.00 and M31-004320.97. Despite the differences in the stars' apparent surface temperatures,  the infrared flux is about one third or more of the total flux of the star.  Many of  the B[e]sgs are
remarkable for a  large  infrared emission  relative to the star, which may
be on the order of one-half the star's total flux or more.
The infrared flux  in M31-004221.78, shown in Figure 5, is equal to the visual 
luminosity\footnote{In this case and some others, there is no data in the 
wavelength range 1--3 $\mu$m where
the total SED normally has a local minimum. We  use simple dust
models to assess their integrated fluxes. A range of dust temperatures probably 
occurs for each object.  We assume: 1) grain emissivity $Q(\lambda)$ is $\propto$ ${\lambda}^{-1}$, 
(2) the mass of dust with temperatures in range $dT$ is a power-law 
$T^{-\alpha} \, dT$ for $T < T_{max}$, and 
(3) $T_{max} = 1200$ K. We estimate the power-law index $\alpha$ by a 
least-squares fit to the observed IR values of $\log {\lambda} f_{\lambda}$. This approach provides accuracies of the order of $\pm$ 15{\%}  for the 
total flux integrated from $\lambda \sim$ 1 to 10 or 22 $\mu$m.}. 
In this figure we also show an 
estimate of the effect of the circumstellar extinction on the star's visual 
SED and our estimate of its temperature. Of course we may have underestimated the star's interstellar extinction, and its
IR flux is doubtful beyond 10 $\mu$m, but the main point is that the star is
nearly enshrouded in dust. More examples can be seen in Figures   
10a and 10b in Paper II. 

The fraction of the total luminosity 
emitted in the infrared is summarized in Table 4 for the 18 B[e] supergiants. 
Two of the B[e]sgs in Table 3, however, show no
evidence for dust out to 20$\mu$m. Although our sample is small, there is some 
correlation with the presence of circumstellar dust and the luminosity of the 
star. The 9  stars with little or no dust have an average luminosity of 4.3 $\times$ 10$^{5}$ L$_{\odot}$ compared with 1.2 $\times$ 10$^{5}$ L$_{\odot}$ for 
the B[e]sgs with dust contributing 25\%  or more of the total luminosity 
of the star.

\subsection{The Spatial Distribution}

When checking the 
emission line stars discussed in this paper for nearby OB stars for the visual
extinction estimates, we noticed that the B[e] supergiants were relatively 
isolated compared with other emission-line stars, the LBVs and LBV candidates.  
\citet{RMH16} showed that the confirmed LBVs in M31 and M33 were found 
primarily in stellar groups.  We have added information about the
stars' spatial identification  with stellar groups such as known H II 
regions and Associations to Tables 2 and 3. In addition to their luminosities 
and temperatures,  and the presence of circumstellar dust, the stars' 
spatial distribution and association with other stellar groups will also 
be relevant to understanding their evolutionary state.

\section{The HR Diagrams}

The HRDs for the M31 and M33 supergiant populations (Figures 1 and 2)
have similar 
distributions. Both show a well-defined upper luminosity boundary or 
Humphreys-Davidson limit, noted in much earlier studies, which is also
clearly visible in Massey's (2016) HR Diagrams. There is a lack of 
luminous O-type stars near the main sequence above the survey magnitude limit, 
shown as a thin dark line on the HR Diagrams. This  is most likely  due to   
a combination of observational selection since they are  more often in 
crowded regions and associated with nebulosity and the shift of their energy 
dsitributions to shorter wavelengths and fainter visual magnitudes. 

Figures 6 and 7 show the LBVs, candidate LBVs, the B[e] supergiants and the
warm hypergiants on their respective HR Diagrams using the temperatures
and derived luminosities in Tables 2 and 3.  For comparison, they are  
over-plotted on the HRDs 
for the supergiant populations which are shown as a faint background.
The confirmed LBVs  
are plotted in their quiescent or hot state, and those that have had a recent 
``LBV eruption'' or 
maximum light stage are illustrated with a dashed line. 
In their quiescent state 
the LBVs lie along the S Doradus or LBV instability strip  
\citep{Wolf89,HD94} also outlined in these Figures. Although the candidate LBVs 
have a wide 
range of temperatures and luminosities, several lie on or near the 
instability strip which adds support to their classification as possible 
candidates.   

\begin{figure}
\figurenum{6}
\epsscale{1.0}
\plotone{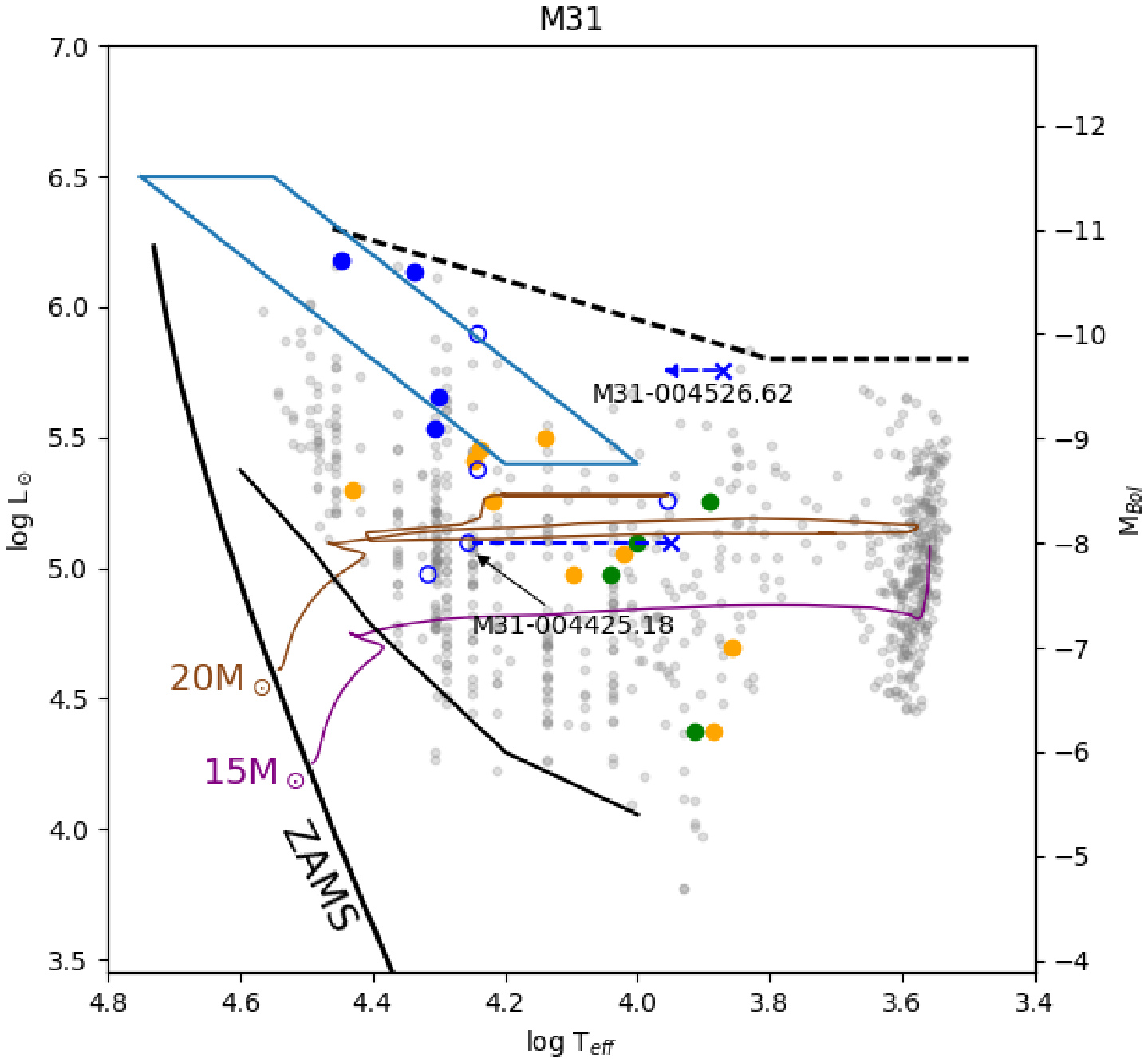}
\caption{A schematic HRD for M31 showing the positions of the confirmed LBVs and
 candidate LBVs shown respectively, as filled and open blue circles, the warm  h
 ypergiants as green circles and the B[e]sgs as orange circles. The LBV transits 
 are shown as dashed lines. The LBV/S Dor instability strip is outlined. The 15 a
 nd 20 M$_{\odot}$ tracks are from  \citet{Ekstrom} with rotation as reference fo
 r the B[e]sgs. (Higher mass tracks are not shown due to crowding.)  The supergia
 nt population from Figure 1 is shown in the background in light gray.}
\end{figure}

\begin{figure}
\figurenum{7}
\epsscale{1.0}
\plotone{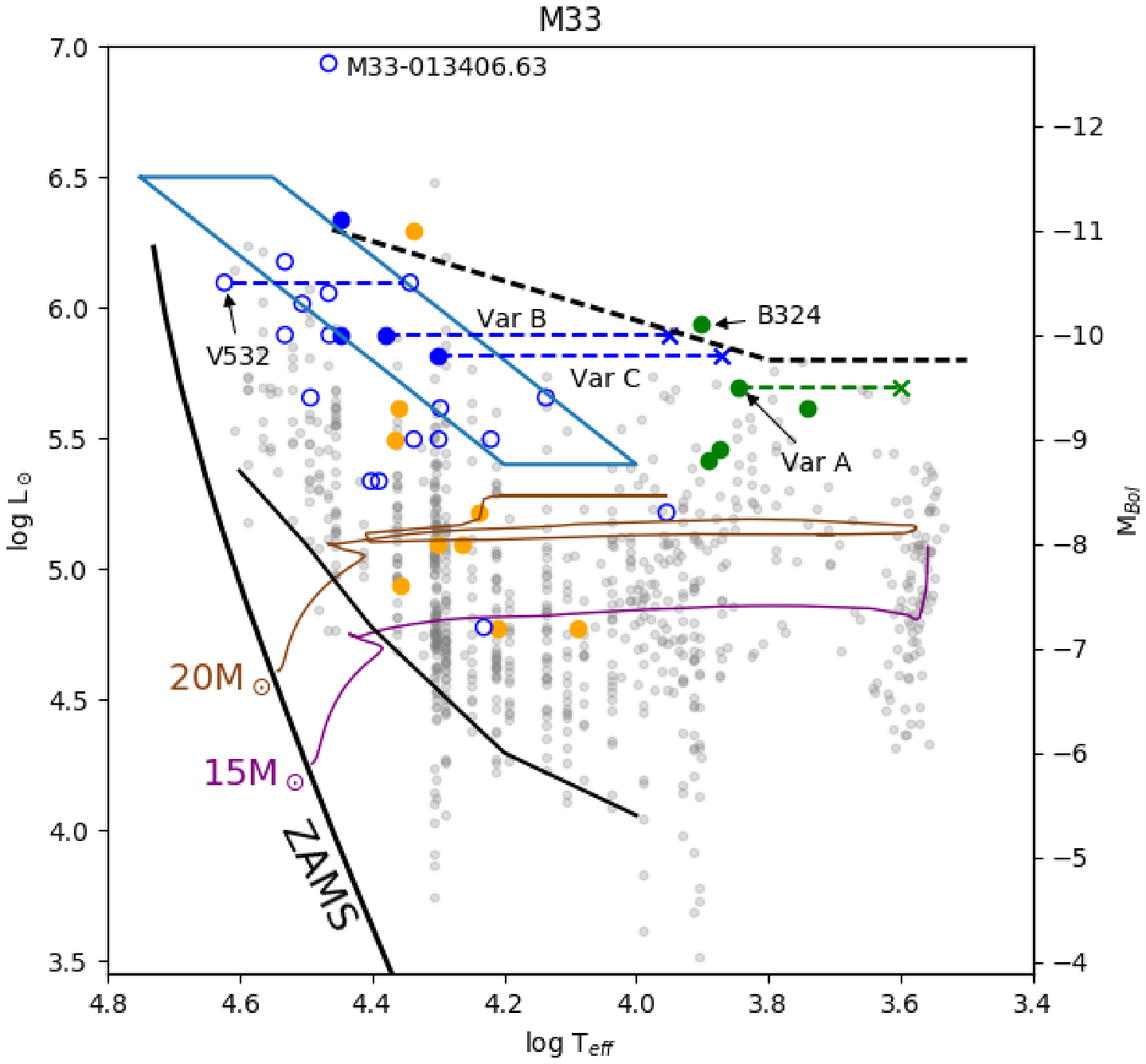}
\caption{A schematic HRD for M33 showing the positions of the confirmed LBVs and
 candidate LBVs shown respectively, as filled and open blue circles, the warm  h
 ypergiants as green circles and the B[e]sgs as orange circles. The LBV transits 
 are shown as dashed lines. The LBV/S Dor instability strip is outlined. The 15 a
 nd 20 M$_{\odot}$ tracks are from  \citet{Ekstrom} with rotation as reference fo
 r the B[e]sgs. (Higher mass tracks are not shown due to crowding.) The supergian
 t population from Figure 2 is shown in the background in light gray.}
\end{figure}

These HR Diagrams illustrate an important point: many normal supergiants 
are found in the LBV instability strip.  Normal mass losing supergiants can 
evolve through the instability strip to lower temperatures. 
LBVs are distinguished from the normal supergiants that occupy the same 
locus in the luminosity-temperature space by their 
proximity to the Eddington limit. They have high L/M ratios, and their 
Eddington factor $\Gamma$ $=$ $L/L_{Edd}$ is high, 0.5 or higher due to
high mass loss events as hot supergiants for stars with initial masses 
$\geq$ 40 -- 50 M$_{\odot}$ or as red supergiants for the less luminous LBVs; see Figure 1 and the discussions in \citet{RMH16} and \citet{KD}. 

In contrast, most of the B[e]sgs in our sample are  less luminous than the LBVs and LBV candidates. Except for one (M33C-15731), they are all below the upper 
luminosity boundary for evolved massive stars, and 
although they have a range in temperatures, their luminosities also place most 
of them  below the LBV instability strip; three in M31 and two in M33 are near the lower bound of the instability strip. There are no very hot stars and the coolest members overlap with the warm hypergiants on the HRDs. 
Published HR Diagrams for the Magellanic Cloud B[e]sgs \citep{Oksala} also show the majority below the upper luminosity boundary for the intermediate temperature supergiants. 

The warm hypergiants  include the 7 stars described in Paper I 
plus two additional 
stars M31-004621.08 (Paper III)  and J013358.05+304539.9 \citep{Kourn17}. We have argued that with their spectroscopic evidence for high mass loss and extensive
circumstellar gas and dust, they are candidates for post main sequence evolution. The one likely exception is B324 in M33 which we suggest may be evolving to
cooler temperatures. Indeed it may be near the end of its redward evolution based on its position on the HR Diagram.  Some of the warm hypergiants also 
spectroscopically resemble the B[e] supergiants. For example,  the strong Ca II 
triplet and [Ca II] doublet in emission  
are also present in some of the  B[e]sgs, and the spectra of two hypergiants, 
M31-0004522.52 and M31-004621.08 also have [Fe II] and [O I] emission (Figure 4 in Paper IV),  the distinguishing spectroscopic characteristic of the 
B[e]sg class.  
Thus the B[e]sgs may be the hotter counterparts of some of the warm hypergiants,
especially the lower luminosity ones in M31. 

\section{Concluding Remarks --- LBVs and B[e] Supergiants}

Numerous authors have suggested possible connections between LBVs and the B[e]sgs
based primarily on their spectroscopic similarities, see \citet{Zick2006} and 
references therein. In paper IV, we discussed their spectroscopic signatures and proposed guidelines for separating the two
classes. The B[e]sgs have forbidden lines of [O I] not observed in LBVs, and in 
many cases Ca II and [Ca II] emission is also present in their spectra,  but not 
in LBVs. 
Equally significant, and a clear distinction between the two, is the presence of
warm dust in most of the B[e]sgs but not in LBVs. Based on their positions
on the HRDs, most of the B[e]sgs appear to originate from  a lower initial mass 
population, 15 -- 40 M$_{\odot}$ (15 -- 30 M$_{\odot}$  for tracks with rotation) compared to the LBVs, most of which are $\geq$ 40 M$_{\odot}$.   

The standard model for the B[e] supergiants is an evolved fast-rotating  
hot supergiant with a
two -component wind -- a fast low density polar wind and slow dense wind in 
the equatorial zone \citep{Zick85,Zick86} with an equatorial disk or ring.
The asymmetric mass loss leads to the formation of the high density disk. 
The extensive dusty region forms in the outerparts of the extended disk protected from the UV radiation from the relatively hot star. Note, however, that  many of the B[e] supergiants not only have dusty ejecta but a significant fraction of their luminosity is radiated in the infrared (Table 4). In more than half of 
the cases with circumstellar dust,  the infrared flux accounts for 25\% or more of the total flux from the star, and as we noted in \S {3.1}, the less luminous B[e]sgs have a larger fraction of their total luminosity due to re-radiation 
by dust.   
Although this dusty circumstellar material is often described as a ``disk'' or even a ``ring'' \citep{deWit}, 
with this much dust that nearly obscures the star in some cases, perhaps ``torus'' is a better  
description for these objects.  \citet{Kastner} also noted the large fraction of the luminosity
in the infrared  and suggested that the equatorial ejecta was a ``puffed up'' or 
flared disk.  

There have long been questions about the variability of the B[e] supergiants and if they are variable, if it is due to variation in the equatorial torus. \citet{Martin}  have monitored these and other emission line stars in M31 and M33, imaging them annually for the past four years with the intention of detecting long-term variability of 0.1 magnitude or greater in BVRI. They find that only one of the 18 stars discussed here is definitely 
variable, and one  other may be variable. Both have a low dust contribution to their luminosities, and none of the very dusty stars showed any significant variability. 
This implies that the variability is driven by changes in the photosphere or material close to the star,  not associated with the dusty material at larger radii. 

Although it is generally agreed that the B[e]sgs are post-main sequence stars, 
their  evolutionary state  is debated with some argung for a post-RSG  or post-yellow/warm supergiant stage \citep{Kastner,Aret} and others for a pre-RSG 
state \citep{Oksala,deWit}. \citet{Oksala} concluded that the  
C$^{12}$/C$^{13}$ ratios measured from 
the CO bandheads in ten  B[e]sgs in the Clouds and the Milky Way are 
consistent with a pre-RSG stage. Only three stars  had the low ratios (5 -14) \citep{Milam} expected for oxygen-rich evolved RSGs and  
post-RSG or YSG stars. 

If the B[e] supergiants are in a pre-red supergiant 
stage,  then, based on their positions on the 
the HR Diagram, we would expect to find them associated with the B-type 
supergiants of similar temperatures.  
The B[e]sgs in our M31 and M33 sample are relatively 
isolated from  nearby OB stars compared to the LBVs and candidates, see \S {3.1} and Table 3, and most are not found in  
the  stellar associations. This difference is  most noticeable for 
the M31 stars. In M33, star forming regions of young stars are found across
its face, so there is a greater possibility of a chance association.
Indeed the B[e]sg  distribution is similar to what we found for 
RSGs in Paper III. Sixty percent of the RSGs lacked nearby OB stars. 
Fifty percent of the B[e]sgs lack nearby stars and 50\% are not 
found in or near a stellar grouping. \citet{Becker} report that none of the ten B[e]sgs in the LMC from Oksala's list have nearby OB stars within the same search 
radius of $\approx$ 11 pc. We find that only three have nearby B-type supergiants which 
is similar to the M31 and M33 sample.

Overall, their spatial distribution supports a more evolved state for most of the 
B[e]sgs. But 
this apparently contradicts the conclusions from the C$^{12}$/C$^{13}$ ratios.  
The more luminous B[e]sgs, such as M33C-15731 and 
those in the Magellanic Clouds above the upper luminosity boundary, 
log L/L$_{\odot}$ $\cong$ 5.8, however,   have not been red supergiants, 
and must have some other origin for their B[e] characteristics. In our study, 
the more luminous are also those with the least dust. Furthermore, 
no measurable CO bandhead emission was detected in those with luminosities 
below 10$^{5}$ L$_{\odot}$ \citep{Oksala}, and the three B[e]sgs with the lower 
C$^{12}$/C$^{13}$ ratios tend to have the lower luminosities. 
Since  the lower luminosity B[e]sgs are also the dustiest with a higher 
fraction of their luminosities in the infrared, these would be the more likely
candidates for post-RSG evolution. Thus the observations, which appear to be 
contradictory, may equally suggest that there is more than one path to become a
B[e] supergiant, based on the spectroscopic criteria. 

If some  of the B[e] supergiants are post-RSGs, then what distinguishes them  from the 
``less luminous'' LBVs that are also likely post-RSG stars? \citet{RMH16} emphasized  that the instability of the LBVs was probably due to their proximity to the 
Eddington  limit for their intial mass; they have an Eddington factor of 
$\approx$ 0.5 or higher, see \citet{HD94}. For the less luminous LBVs, this is attributed to 
mass loss as a red supergiant. The same may be true for the B[e]sgs. Indeed,
their Eddington factor would be enhanced by rapid rotation. 
Binarity could be a factor; \citet{deWit} suggested that some of the Galactic B[e] stars are short-period binaries with circumbinary disks. This does not necessarily 
mean that  LBVs are binaries. Their instability is most likely a surface 
phenomenon, while in the B[e]sgs, the proximity of a companion to a rapidly 
rotating star may generate the mass loss in the equatorial region without the 
LBV eruption.  We also note that many of the B[e] supergiants lie below the 
LBV instability strip which suggests that the B[e]sg stage may be a more
common end state for lower mass supergiants, 15 - 20 M$_{\odot}$.

In summary, the LBVs and B[e]sgs are not related in an evolutionary sense. There are sufficient differences in their spectra, their luminosities,
spatial distribution and the presence or lack of circumstellar dust, to conclude 
that one 
group does not evolve into the other.  We support the  conclusions in our previous papers \citep{HD94,RMH13,RMH16}  that the ``less luminous'' LBVs and  the warm hypergiants are post-RSGs. Here we  suggest that some of  the B[e] supergiants may also 
be  examples of post-red supergiant evolution.

\acknowledgements
Research by R. Humphreys and K. Davidson on massive stars is supported by  
the National Science Foundation AST-1109394. J. C. Martin's collaborative work is supported by NSF grant AST- 1108890. We thank Alexander Becker and Dominik Bomans for 
communicating their resutls on the spatial distribution of the B[e] supergiants in the 
LMC.

\appendix 
\section{Two Exceptionally Luminous O-type Supergiants in M31}

Two of the supergiants in M31 with very high extinction values, $>$ 3 mag, are 
classed as isolated (I) by \citet{Massey16} which means that their spectra and photometry are not 
contaminated by unresolved nearby stars. It does not mean that they are physically isolated from other luminous stars.  We checked their positions and confirm that neither of these stars is blended with nearby stars. Their extinction values derived from their observed colors however lead to  
luminosities $\sim$ 10$^{7}$ L$_{\odot}$.  

{\it J004246.85+413336.4} (O3-O5 If) has an M$_{v}$ of -9.8 mag and M$_{Bol}$ of -13.5 mag (10$^{7.3}$ L$_{\odot}$) 
with its A$_{v}$ of 3.17 mag. It is in a small group of what are apparently hot stars based on their colors.  Although it is not in a designated  association, it is in a prominent H II region \citep{Azimlu} and a  dust cloud, D392, cataloged by \citet{Hodge}, which of course accounts for the high local interstellar extinction. The mean visual extinction from four nearby stars 
is $\approx$ 3.0 mag and from the neutral hydrogen density it is 1.8 mag with a possible maximum  value of  3.1 mag, see \S {3}.  Thus the high extinction is confirmed from nearby stars and from the star's location in a dust cloud. 

{\it J004158.87+405316.7} (O9.5 I) is isolated with only one nearby star. Like the prevous star, it is not in a known association, but is on the edge of a cataloged dust cloud, D222 \citep{Hodge}, and there is no associated nebulosity.  Its  observed photometry yields 
an M$_{v}$ of -9.2 and M$_{Bol}$ of -12.0 mag (10$^{6.7}$ L$_{\odot}$) with A$_{v}$ of 3.25 mag placing it way above the hot supergiants of comparable spectral types on the HR Diagram. The only nearby star has a lower A$_{v}$ of 1.8 mag and derived colors appropriate
to an early B-type supergiant. The difference in the extinction values
may be due to a gradient in the dust near the edge of the cloud. The  extinction from the N$_{HI}$ density is 0.9 mag  with a maximum of 1.5 mag similar to the nearby star.  

Neither star is on archived ``HST mages'', so it is not possible to check for possible
blending with unresolved nearby  stars. 
We suspect that both stars are most likely more than one star unresolved at their distance, perhaps similar to J013406.63+304147.8 (UIT 301, B416) in M33, Paper II, although \citet{Crowther} has identified WN-type stars with luminosities of 10$^{6.5}$ to 10$^{6.9}$ L$_{\odot}$ in R136a in the LMC. These two stars in M31 however are relatively isolated by comparison with R136a. Both deserve closer scrutiny with higher resolution spectra and imaging.


\begin{deluxetable}{lccccc}
\tablewidth{0 pt}
\tabletypesize{\footnotesize}
\tablenum{1} 
\tablecaption{Number of Stars in Different Spectral Type Groups} 
\tablehead{
\colhead{Galaxy} &
\colhead{OB stars} &
\colhead{A-type supergiants}  &
\colhead{YSGs} &
\colhead{RSGs} &
\colhead{Total} 
}
\startdata
M31     &  333   & 52   &  50   & 295    &  730     \\
M33     &  516   & 127   & 57   &  101   &  801     \\  
\enddata
\end{deluxetable}

\begin{deluxetable}{lccccccc}
\tablewidth{0 pt}
\rotate 
\tabletypesize{\scriptsize}
\tablenum{2} 
\tablecaption{Extinction, Luminosities, and Temperatures - LBVs and candidates}  
\tablehead{
\colhead{Star} &
\colhead{A$_{v}$ (stars)} &
\colhead{A$_{v}$ (H I)}  &
\colhead{Adopted A$_{v}$} & 
\colhead{M$_{v}$} &
\colhead{T}  & 
\colhead{M$_{Bol}$} &  
\colhead{Stellar Group\tablenotemark{a}} \\
    &   (mag)  & (mag)  &  (mag)  & (mag) &  $\arcdeg$K & (mag) &  }
\startdata 
      &        &      &  M31  &   &  &   & \\
LBVs  &        &      &       &   &   &   & \\ 
AE And  & 0.9(1)   & 0.9   & 0.9  & -7.9: & 20000:   & -9.4:  &  A170, H II\\
AF And  &   \nodata    &  1.1   &  1.1 &  -8.2:  & 28000: &  -10.7:& \nodata \\
Var A-1\tablenotemark{b} &  1.4(1)   &  1.9 & 1.4  & -8.7  & 21700  & -10.6: &  A42 \\ 
Var 15\tablenotemark{b}    &  \nodata  &  1.3  & 1.3  &  -7.3:  & 20200: & -9.1: & A38  \\
M31-004526.62\tablenotemark{c} &  1.6(1) & 1.3   & 1.5  & \nodata & \nodata &  -9.65  &  A45, H II  \\
   &        &         &     &   &  &  &  \\
LBV Candidates  &    &    &     &   &    &   \\ 
M31-003910.85   &  1.0(3)  &  1.0  &  1.0 & -7.2 & 17400 & -8.7  & A127 \\
M31-004051.59\tablenotemark{d}   &  1.7  & 1.6  &  0.6 & -8.1 &  9000 & -8.4  & A82 \\
M31-004411.35   &  2.3(1): &  0.8  &  2.3: & -8.6 & 17400  &  -10   & A10 \\
M31-004425.18\tablenotemark{e}   &  \nodata &  0.7  &  0.7 &      & 18000 &  -8   & isolated(A9)\\
M31-004444.00   &  0.5(3)  &  1.0  & 0.5  & -5.9 & 20700 &  -7.7 &  A54\\
   &        &         &     &   &  &  &  \\
    &        &       &   M33  &   &  &  &  \\
LBVs  &        &      &       &   &   &   & \\  
Var B   &  0.4  & 0.3 & 0.4  & -7.7  & 24000:  &  -10.0 &  A142\\ 
Var C\tablenotemark{f}  & 0.5(2)     & 0.4 &  0.5  &  -8.6 & 20000:  &  -9.8 &  H II  \\
Var 83  & 0.8(2)   & 0.9 &  0.8  &  -8.8: & 28000:  & -11.1: &   A101,A103, H II \\  
Var 2  &  0.9(1)  &  1.0 &  0.9 &  -7.2 &  28000:  & -10.0:  &  A100  \\   
   &        &         &     &   &  &  &  \\
LBV Candidates  &    &    &     &   &    &   \\
M33C-2976  &  0.4(2) & 0.5 &  0.4 &  -5.9 &  17000 &  -7.2 &  A124 \\
M33C-4174  &  0.8(3) & 0.9 & 0.8  &  -7.3 &  19900 &  -9.0 & A130 \\
UIT008\tablenotemark{g}  &  0.4(1) & 0.8 & 0.6  &  -7.5 &  34000 & -10.7  & H II, A27 \\
M33C-14239 &  0.3(2) & 0.4 & 0.4  &  -7.6 &  16600 &  -9.0 &  Spiral arm \\
M33C-4640\tablenotemark{h} & 0.6(2) & 0.6 &  0.6 & -8.1 &  9000 & -8.3 &  A128\\
M33C-25255 &  0.7(3) & 0.6 & 0.6  & -6.3 & 24500 & -8.6 & A137\\
M33-013317.22 &  0.7(2) & 0.5 & 0.7 &  -6.5 &  29100 & -10.0 & A17 \\
M31-013334.06 &  \nodata & 0.6 & 0.6 & -7.6 &  19800 & -9.3  & \nodata \\
M33C-7024  & 0.3(3)  &  0.3 & 0.3 & -6.3 &  25200 & -8.6  & Spiral Arm \\
M33C-15235 & 1.1(4)  &  0.8 & 1.0 & -7.8 &  29200 & -10.4 & A64 \\
M33C-5916  & 0.7 (1) &  0.6 & 0.7 &  -6.9 & 34000 &  -10.0 & A6 \\
M33-013406.63\tablenotemark{i} & 0.8(1) & 0.7  & 0.7:  & -9.1: & 29200: & -12.6: & H II \\
M33C-21386 & 0.9(4)  & 1.0 &  0.9 & -8.2  & \nodata & \nodata & A71 \\ 
M33C-10788 & 0.8(3)  & 0.5 &  0.8 & -7.3  & 32000 & -10.3 & A100 \\
M33-013424.78 &  \nodata & 0.9 & 0.9 & -8.8 & 13700 & -9.4 & A102 \\
M33C-20109    &   2.4(1):   & \nodata  &  2.4:  & -8.5   & \nodata  & \nodata    &    \\
M33-013432.73 & 0.6(2) & 0.6 & 0.6 & -6.0 & 21700 & -9.0 & A84, N604 \\
M33C-16364  &  0.4(6)  & 0.5 &  0.4 & -6.7 & 31100 & -9.4 & A88 \\
V532/GR290\tablenotemark{j}  & \nodata  & 0.6 &  0.6 &  var & 42000-22000 & -10.4-10.7 & A89:  \\ 
\enddata
\tablenotetext{a}{References: \citet{Hodge}, \citet{HS80}} 
\tablenotetext{b}{Var A-1 and Var 15 have both shown spectroscopic and photometric variability in the past few years (see Paper IV). Their SEDs are uncertain.}
\tablenotetext{c}{New LBV in M31. See \citet{RMH15}. Its quiescent or minimum light state is uncertain.}
\tablenotetext{d}{See Papers III and IV, LBV candidate or post-RSG. The adopted A$_{v}$ is from the observed colors. A$_{v}$ from the nearby stars would yield M$_{v}$ $= -9.2$, M$_{Bol}$ $= -9.5$.} 
\tablenotetext{e}{A probable low-luminosity LBV. See discussions in Papers II and IV.} 
\tablenotetext{f}{Var C is in eruption. See \citet{RMH2014}}
\tablenotetext{g}{UIT 008 (J013245.41+303858.3). See Papers II and IV.}   
\tablenotetext{h}{M33C-4640 may be a post-RSG star, not an LBV candidate (Paper II), but it has weak Fe II emission.}
\tablenotetext{i}{UIT 301, B416) This very lumninous hot supergiant is
very likely more than one star as discussed in Paper II. It is listed as an LBV candidate because of the [Fe II] and Fe II emision in its spectrum.}
\tablenotetext{j}{Romano's star, see discussion in Paper II.}
\end{deluxetable}

\begin{deluxetable}{lccccccc}
\tablewidth{0 pt}
\rotate 
\tabletypesize{\scriptsize}
\tablenum{3} 
\tablecaption{Extinction, Luminosities, and Temperatures - B[e] Supergiants and Warm Hypergiants}  
\tablehead{
\colhead{Star} &
\colhead{A$_{v}$ (stars)} &
\colhead{A$_{v}$ (H I)}  &
\colhead{Adopted A$_{v}$} & 
\colhead{M$_{v}$} &
\colhead{T}  & 
\colhead{M$_{Bol}$} &  
\colhead{Stellar Group\tablenotemark{a}} \\
    &   (mag)  & (mag)  &  (mag)  & (mag) &  $\arcdeg$K & (mag) &  }
\startdata 
      &        &      &  M31  &   &  &   & \\
B[e] supergiants  &        &      &       &   &   &   & \\ 
M31-004043.10*\tablenotemark{b}  & 1.5(1)   & 0.8   & 1.0  & -6.8  & 16500   & -8.4  &  A120\\
M31-004057.03 &   \nodata  &  0.7   &  0.7 &  -6.3 & 7700: & -6.2: & \nodata \\
M31-004220.31* &  \nodata   &  1.1 & 1.1  & -6.6  & 12500 &  -7.7 & \nodata \\ 
M31-004221.78*  &  \nodata  &  0.95  & 1.0  &  -5.8 & 7200: & -7.0 & \nodata  \\
M31-004229.87* & 1.4(2) & 2.1   & 1.4  &  -7.0  &  17600 &  -8.8  & \nodata \\
M31-004320.97* & \nodata & 1.5  & 1.5  &  -6.7  &  10500 & -7.9 & \nodata \\
M31-004415.00* &  \nodata & 0.9 &  0.9 &  -7.0  &  17300 & -8.9 & \nodata\\
M31-004417.10* &  0.7(1)  & 1.7 &  0.7 &  -8.0  &  13800 &  -9.0 & A32:\\
M31-004442.28* &  1.2(1)  & 1.7 &  1.2 &  -5.9  & 26900  &  -8.6 & \nodata\\
   &        &         &     &   &  &  &  \\
Warm Hypergiants  &    &    &     &   &    &   \\ 
M31-004322.50* &  1.5(1)  &  1.1 &  1.5 & -5.6 & 8200  & -6.2  & Spiral Arm \\
M31-004444.52*   &  1.5(3)  & 1.3  &  1.5 & -7.8 & 6600 & -8.4  & A41 \\
M31-004522.58*   &  1.2(1) &  1.1 &  1.2 & -7.1 & 11000  & -7.7   & A46 \\
M31-004621.08*   &  1.2(1) & 0.3   & 1.2 & -7.6 & 10000 &  -8.0   & A104:\\
   &        &         &     &   &  &  &  \\
    &        &       &   M33  &   &  &  &  \\
B[e] supergiants  &        &      &       &   &   &   & \\  
M33-013242.26    &  0.7(2) & 0.8 & 0.7 & -7.8 & 22900: & -9.3 & A118 \\
M33-013324.62*    &  \nodata & 0.4 & 0.4 & -5.3 & 22800 & -7.6 &   \nodata \\
M33C-7256*     & 0.5(3)  & 0.3  & 0.5 & -5.6 & 18400 & -8.0 & A14 \\
M33C-6448     & 0.5(2)  & 0.8  & 0.5 & -6.9 & 17300 & -8.3 & A9 \\
M33C-24812*    & \nodata & 0.7  & 0.7 & -6.2 & 20000 & -8.0 & A35 \\
M33C-15731*    & 1.2(2)  & 0.7  & 1.2 & -8.9 & 21700 & -11.0: & A64(bulge) \\
M33-013426.11*  & \nodata & 0.6  & 0.6 & -6.1 & 12300 & -7.2 & \nodata \\
M33-013459.47*    & \nodata & 0.7  & 0.7 & -6.8 & 23200 & -9.0 & \nodata \\
M33-013500.30*    & 0.3(3)  & 0.7  & 0.3 & -5.5 & 16200 & -7.2 & A88 \\
   &        &         &     &   &  &  &  \\
Warm Hypergiants  &    &    &     &   &    &   \\
Var A\tablenotemark{c}* &  0.6-0.9  & 1.0 & \nodata & \nodata & 7000: & -9.5  & A130 \\
B324\tablenotemark{d}  & 0.9(3) & 0.8  &  0.8 &  -10.2 & 8000  & -10.1 & A67  \\
N093351\tablenotemark{e}* &  0.45(2) & 0.8  &  0.45 & -8.8 &  7800 & -8.8  & A142 \\
N125093\tablenotemark{e}* &  0.9(1)  & 0.8  &  1.6  & -8.8 &  7500 & -8.9  & A105 \\ 
J013358.05+304539.9\tablenotemark{f}* & 0.8(2) & 0.6 & 0.8 & -9.2 & 5000-6000 & -9.3: & A68 \\
\enddata
\tablenotetext{a}{References: \citet{Hodge}, \citet{HS80}} 
\tablenotetext{b}{* -- circumstellar dust}
\tablenotetext{c}{Var A is optically obscured. See \citet{RMH87,RMH06} for a discussion of its light curve, spectrum and energy distribution.}
\tablenotetext{d}{B324 is possibly the visually most luminous star in M33. See Paper II for a discussion of its spectrum and SED. B324 is probably not a post-RSG, but more likely evolving to cooler temperatures.}
\tablenotetext{e}{N is the designator from the \citet{Valeev} survey used in Papers I and II. In Papers III and IV we used a V followed by the number.}
\tablenotetext{f}{See \citet{Kourn17}}
\end{deluxetable}

\begin{deluxetable}{lccc}
\tablewidth{0 pt}
\tabletypesize{\footnotesize}
\tablenum{4} 
\tablecaption{B[e] Supergiant Luminosities } 
\tablehead{
\colhead{Star} &
\colhead{Visual L$_{\odot}$}  &
\colhead{Infrared L$_{\odot}$} &
\colhead{Ratio (IR/Total} 
}
\startdata
    &  M31   &    &         \\
M31-004043.10 & 1.6 $\times 10^{5}$ & 1.8 $\times 10^{4}$ &  0.10 \\
M31-004057.08\tablenotemark{a} & 1.8$\times 10^{4}$ &  \nodata & \nodata \\
M31-004220.31 & 6.5 $\times 10^{4}$ & 1.9 $\times 10^{4}$ & 0.23 \\ 
M31-004221.78 & 2.2 $\times 10^{4}$ & 2.8 $\times 10^{4}$ & 0.55 \\
M31-004229.87 & 1.8 $\times 10^{5}$ & 6.8 $\times 10^{4}$ & 0.27 \\
M31-004320.97 & 6.7 $\times 10^{4}$ & 5.0 $\times 10^{4}$ & 0.43 \\
M31-004415.00 & 1.6 $\times 10^{5}$ & 8.4 $\times 10^{4}$ & 0.34 \\
M31-004417.10 & 3.0 $\times 10^{5}$ & 1.4 $\times 10^{4}$ & 0.04 \\
M31-004442.28 & 2.0 $\times 10^{5}$ & 1.1 $\times 10^{4}$ & 0.01 \\
    &        &    &        \\
    &  M33   &    &         \\  
M33-013242.26 & 4.0 $\times 10^{5}$ & \nodata & \nodata \\ 
M33-013324.62 & 7.1 $\times 10^{4}$ &  1.3 $\times 10^{4}$ &  0.15 \\
M33C-6448     & 1.7 $\times 10^{5}$ & \nodata & \nodata \\ 
M33C-7256     & 6.5 $\times 10^{4}$ &  6.1 $\times 10^{4}$ &  0.48 \\
M33C-24812    & 1.2 $\times 10^{5}$ &  7.0 $\times 10^{3}$ &  0.05 \\
M33C-15731    & 2.0 $\times 10^{6}$ &  1.0 $\times 10^{5}$ &  0.05  \\
M33-013426.11 & 4.3 $\times 10^{4}$ &  1.5 $\times 10^{4}$ &  0.26 \\
M33-013459.47 & 3.2 $\times 10^{5}$ &  1.4 $\times 10^{4}$ &  0.04 \\
M33-013500.30 & 4.0 $\times 10^{4}$ &  1.9 $\times 10^{4}$ & 0.32 \\
\enddata
\tablenotetext{a}{The star is very faint. No 2MASS, IRAC or WISE photometry.}
\end{deluxetable}




\begin{figure}
\figurenum{3}
\epsscale{1.0}
\plotone{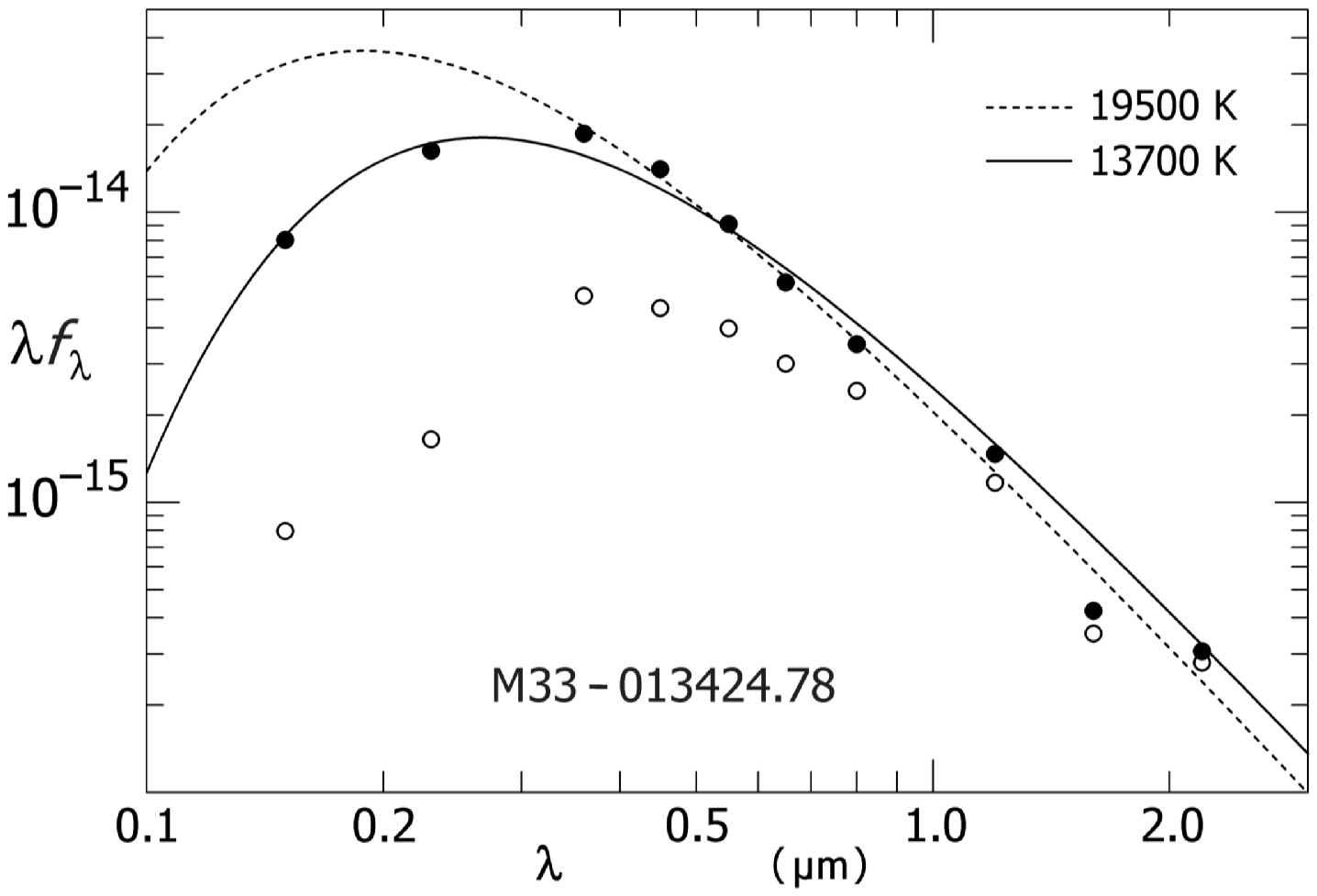}
\caption{The SED (Watts/m$^{2}$ vs wavelength in microns) for the LBV candidate M33-013424.78 illustrating the constraint 
on the Planck fit and estimated temperature provided by the FUV and NUV fluxes. The solid curve with the UV fluxes yields a temperature of 13700 K compared with a
higher temperature of 19500 K without the UV; this difference may be due to the Balmer edge near 
0.36$\mu$m.  The extinction-corrected magnitudes are shown as filled circles and the observed magnitudes  as open circles.   
}
\end{figure}

\begin{figure}
\figurenum{4}
\epsscale{0.7}
\plotone{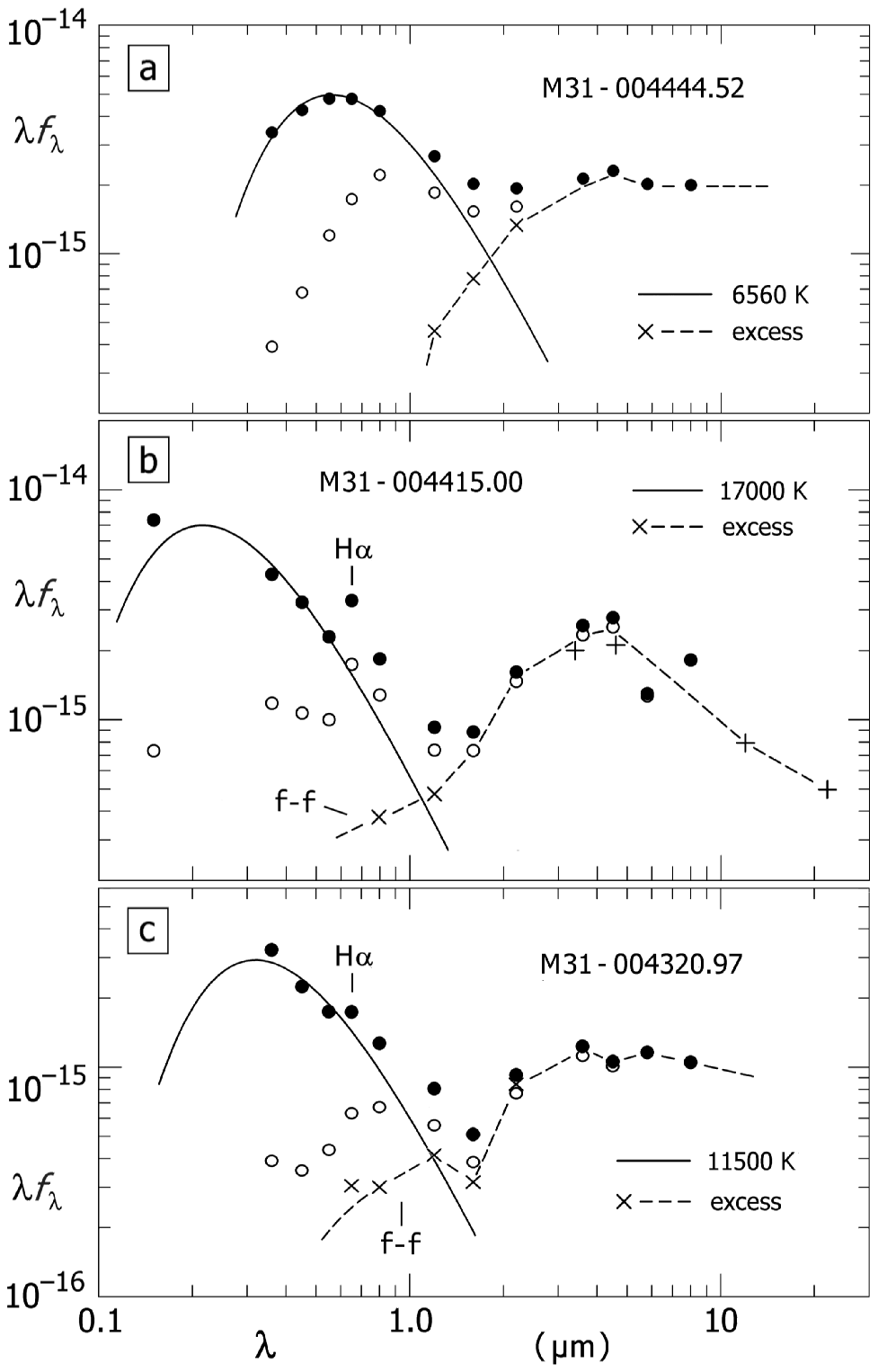}
\caption{The SED (Watts/m$^{2}$ vs wavelength in microns) for the warm hypergiant M31-004444.52 (a), and two B[e] supergiants, M31-004415.00 (b) and M31-004320.97 (c). The extinction-corrected magnitudes are shown as filled circles and the observed magnitudes  as open circles. The IR excess is shown as a dashed line. Both of the B[e]sgs have free-free emision in addition which contributes to the flux longwards of H$\alpha$ and in the near-IR.}
\end{figure}

\begin{figure}
\figurenum{5}
\epsscale{1.0}
\plotone{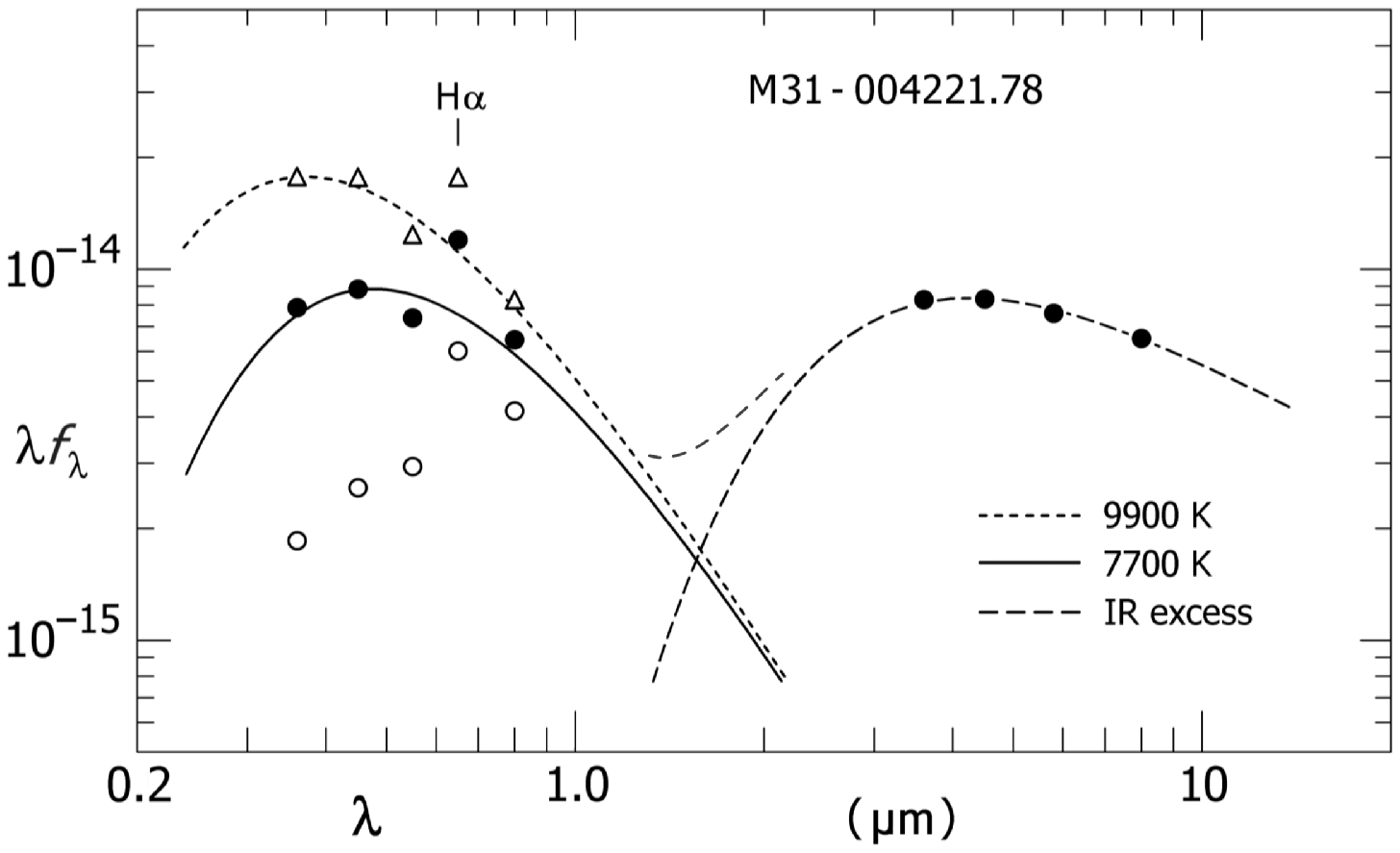}
\caption{The SED (Watts/m$^{2}$ vs wavelength in microns) for the B[e]sg 
M31-04221.78. The infrared flux contributes more than half of this star's total flux.  The open circles are the observed visual photometry. Note the strength 
of H$\alpha$. The visual magnitudes corrected for interstellar extinction and the mid-IR fluxes are shown as filled circles. The  open triangles and short-dashed Planck curve illustrate an estimate of the effect of the additional extinction from the circumstellar dust on the star's temperature. }
\end{figure}




\begin{thebibliography}{}
\bibitem[Aret et al.(2012)]{Aret}Aret, A., Kraus, M., Muratore, M. F. \& Ferna
ndes, M. B. 2012, \mnras, 423,284
\bibitem[Azimlu et al.(2011)]{Azimlu}Azimlu, M., Marciniak, R., \& Barmby, P. 2011, \aj, 142, 139  
\bibitem[Becker et al.(2017)]{Becker}Becker, A., Bomans, D.~J., et al, 2017, in preparation 
\bibitem[Braun et al.(2009)]{Braun}Braun, R., Thilker, D. A., Walterbos, R. A. 
M. \&  Corbelli, E, 2009, \apj, 695, 937
\bibitem[Cardelli et al.(1989)]{Cardelli}Cardelli, J. A., Clayton, G. C., \& Mathis, J. S. 1989, \apj, 345, 245 
\bibitem[Crowther et al.(2010)]{Crowther}Crowther, P. A. et al. 2010, \mnras, 408, 731,
\bibitem[Davidson et al.(2016)]{KD}Davidson, K., Humphreys, R.M, \& Weis, K. 2016, arXiv:1608.02007(v2)
\bibitem[de Wit et al.(2014)]{deWit}de Wit, W. J., Oudmaijer, R. D. \& Vink, J. S. 2014, Advances in Astronomy, 2014, id.270848  
\bibitem[Flower(1996)]{Flower}Flower, P. J. 1996, \apj, 469, 355
\bibitem[Ekstrom et al.(2012)]{Ekstrom}Ekstrom, S. et al. 2012, \aap, 537, A146 
\bibitem[Gordon et al.(2003)]{KGordon}Gordon, K. D., Clayton, Geoffrey C., Misselt, K. A., Landolt, A.  U. \& Wolff, M. J. 2003, \apj, 594, 279  
\bibitem[Gordon et al.(2016)]{Gordon}Gordon, M. S., Humphreys, R. M. \& Jones, T
. J. 2016, \apj, 825, 50 (Paper III)
\bibitem[Gratier et al.(2010)]{Gratier}Gratier, P. et al., 2010, \aap, 522A, 3G
\bibitem[Hodge(1981)]{Hodge}Hodge, P. W., Atlas of the Andromeda Galaxy, 1981, University of Washington Press, (Seattle and London)
\bibitem[Humphreys \& Sandage(1980)]{HS80}Humphreys, R. M \& Sandage, A. 1980, \apjs, 44, 319
\bibitem[Humphreys et al.(1987)]{RMH87}Humphreys, R. M., Jones, T. J. \& Gehrz, 
R. D. 1987, \aj, 94, 315 
\bibitem[Humphreys \& Davidson(1994)]{HD94}Humphreys. R. M. and Davidson, K. 1994, \pasp, 106, 1025 
\bibitem[Humphreys et al.(2006)]{RMH06}Humphreys, R. M., Jones, T. J., Polomski,
 E., et al. 2006, \apj, 131, 2105  
\bibitem[Humphreys et al.(2013a)]{RMH13}Humphreys. R. M., Davidson, K, Grammer, S., Kneeland, N., Martin, J. C., Weis, K. \& Burggraf, B. 2013a, \apj, 773, 46 (Paper I)  
\bibitem[Humphreys et al.(2014a)]{RMH2014}Humphreys, R. M., Davidson, K., Gordon, M. Weis, K. Burggraf, B., Bomans, D.~J. \& Martin, J.~C. 2014a, \apjl, 782L, 21H  
\bibitem[Humphreys et al.(2014b)]{RMH14}Humphreys, R. M.,Weis, K.,Davidson, K
., Bomans, D.~J., \& Burggraf, B. 2014b, \apj, 790, 48 (Paper II) 
\bibitem[Humphreys et al.(2015)]{RMH15}Humphreys, R. M., Martin, J. C., \& Gordon, M. S., 2015, \pasp, 127, 347
\bibitem[Humphreys et al.(2016)]{RMH16}Humphreys, R. M.., Weis, K.,  Davidson, K., \& Gordon, M. S. 2016, \apj, 825, 64  
\bibitem[Humphreys et al.(2017)]{RMH17}Humphreys, R. M.., Gordon, M. S., Martin, J. C., Weis, K., \& Hahn, D. 2017, \apj, 836, 1 (Paper IV)  
\bibitem[Jennings et al.(2014)]{Jennings} Jennings, Z. G., Williams, B. F., Murphy, J. W.,  et al. 2014, ApJ., 795, 170 
\bibitem[Kastner et al.(2010)]{Kastner}Kastner, J. H., Buchanan, C., Sahai, R., Forrest, W. J., \& Sargent, B. A. 2010, \aj, 139, 1993 
\bibitem[Knapp et al.(1973)]{Knapp}Knapp, G. R., Kerr, F. J. \& Rose, W. K. 1973, \apjl, 14, 187
\bibitem[Kourniotis et al.(2017)]{Kourn17}Kourniotis, M, Bonanos, A. Z., Yuan, W. M., Macri,L. M., Garcia-Alvarez, D. \& Lee, C.-H. 2017, \aap, 601, 76  
\bibitem[Martin \& Humphreys(2017)]{Martin}Martin, J. C. \& Humphreys, R. M. 2017, submitted to \aj 
\bibitem[Martins et al.(2005)]{Martins}Martins, F., Schaerer, D. \& Hillier, D. J. 2005, \aap, 436, 1049 
\bibitem[Massey et al.(1996)]{Massey96}Massey, P., Bianchi, L., Hutchings, J. B., \& Stecher, T. P. 1996, \apj, 469, 629 
\bibitem[Massey et al.(2006)]{Massey06}Massey, P., Olsen, K. A. G., Hodge, P. W. et al. 2006a, \aj, 131, 2478   
\bibitem[Massey et al.(2016)]{Massey16}Massey, P., Neugent, K. F., \& Smart, B. M. 2016 \aj, 152, 62
\bibitem[Milam et al.(2009)]{Milam}Milam, S. N., Woolf, N. J. \& Ziurys, L. M. 2009, \apj, 690, 837 
\bibitem[Mudd \& Stanek(2015)]{Mudd}Mudd, D. \& Stanek, K. Z. 2015, \mnras, 450, 3811
\bibitem[Oksala et al.(2013)]{Oksala}Oksala, M. E., Kraus, M., Cidale, L. S., Mu
ratore, M. F., \& Borges Fernandes, 2013, \aap, 558, A17
\bibitem[Riess et al.(2012)]{M31Ceph}Riess, A. G., Fliri, J., \& Valls-Gabaud, D . 2012, \apj, 745, 156
\bibitem[Savage \&  Jenkins(1972)]{Savage}Savage, B. D. \& Jenkins, E. B, 1972, \apj, 174, 491 
\bibitem[Scowcroft et al.(2009)]{M33Ceph}Scowcroft, V., Bersier, D., Mould, J. R ., \& Wood, P. R. 2009, \mnras, 396, 1287
\bibitem[Smartt et al.(2009)]{Smartt09}Smartt, S.~J., Eldridge, J.~J., Crockett, R.~M., \& Maund, J.~R. 2009, \mnras, 395, 1409
\bibitem[Smartt et al. (2015)]{Smartt15}Smartt,S.~J.2015, \pasa, 32, 16
 et al. 1995, \aap, 314, 131 
\bibitem[Van Dyk(2005)]{VanDyk05}Van Dyk, S. D. 2005, The Fate of the Most Massive Stars, ASP Conf. Ser 332, (ed. R. M. Humphreys \& K. Z. Stanek, Astron. Soc. Pacific, San Francisco), 47  
\bibitem[Van Dyk and Matheson(2012)]{vandyk}Van Dyk, S. D. and Matheson, T. 2012, in Eta Carinae  and the Supernova Impostors, Astrophys.\ \& Sp.\ Sci.\ Library 384 (ed.\ K.\ Davidson \& R.M.\ Humphreys, Springer Media, New York),  249 
\bibitem[Valeev et al.(2010)]{Valeev}Valeev, A. F., Sholukhova, O. N., \& Fabrik
a, S. N. 2010, Astrophysical Bull., 65, 140 
\bibitem[Wolf(1989)]{Wolf89}Wolf, B. 1989, \aap, 217, 87
\bibitem[Zickgraf et al.(1985)]{Zick85}Zickgraf, F.-J., Wolf, B., Stahl, O., Leitherer, C., \& Klare, G. 1985, \aap, 143, 421 
\bibitem[Zickgraf et al.(1986)]{Zick86}Zickgraf, F.-J., Wolf, B., Leitherer, C.,
Appenzeller, I., \& Stahl, O. 1986, \aap, 163, 119 
\bibitem[Zickgraf(2006)]{Zick2006}Zickgraf, F.-J. Stars with the B[e] Phenomenon,
ASP Conf. Ser 355, (ed. M Kraus \& A. S. Miroshnichenko), 211
\end{thebibliography}
\end{document}